\documentclass[aps,prd,twocolumn,nofootinbib,showpacs,showkeys]{revtex4-1}

\usepackage{graphicx,color}
\usepackage{enumerate}
\usepackage{amsmath,amssymb,amsthm}
\usepackage[paperwidth=210mm,paperheight=297mm,centering,hmargin=20mm,vmargin=20mm]{geometry}
\usepackage{float}
\usepackage{mathrsfs}
\usepackage{hyperref}
\usepackage{physics}
\usepackage[T1]{fontenc}
\usepackage{slashed}
\usepackage[usenames,dvipsnames]{xcolor}
    \definecolor{officegreen}{rgb}{0.0, 0.5, 0.0}

\usepackage{accents}
\newcommand*{\dt}[1]{%
  \accentset{\mbox{\large\bfseries .}}{#1}}

\renewcommand{\i}{\text{i}}

\newcommand{\e}{\text{e}}

\newcommand{\be}{\begin{equation}}
\newcommand{\ee}{\end{equation}}
\newcommand{\beq}{\begin{eqnarray}}
\newcommand{\eeq}{\end{eqnarray}}

\date{\today}

\begin{document}
\title{Effects of a Quantum or Classical Scalar Field on the Entanglement Entropy of a Pair of Universes}

\author{Samuel Barroso Bellido}
\email{samuel.barroso-bellido@usz.edu.pl}
\affiliation{Institute of Physics, University of Szczecin, Wielkopolska 15, 70-451 Szczecin, Poland}

\begin{abstract}
Using the formalism of the third quantization in canonical quantum gravity, the entropy of entanglement of a pair of universes created in the multiverse from the vacuum has lately been calculated. Here, we examine the differences between considering a scalar field as quantum or classical on the entanglement entropy of three different pairs: de-Sitter universes, flat stiff matter dominated universes, and closed universes with a scalar field.  We find that the entanglement entropy is unchanged, zero, or dependent on the treatment of the scalar field, respectively. 
\end{abstract}

\maketitle

\section{Two Distinct Wheeler-DeWitt Equations}

One of the weak points of Canonical Quantum Gravity (CQG) \cite{DEWITT} is the assumption that the ADM variables \cite{ADM} can be quantized \`{a} la Dirac without an experimental reason. This also happens when we quantize the matter fields of the minisuperspace. Yet, it may be perfectly admissible.  Some studies has been done about the difference of the classical and the quantum dynamic of different models \cite{Bianchi, Vakili} where different behaviors have been found for both considerations.  However, when a quantum description of gravity is deployed,  two choices emerge related to the scalar fields. Taking into account scalar fields for FLRW universes, there are two different ways to proceed: to keep the scalar field as a quantum variable, or to treat it as classical fulfilling some equation of state. The quantum treatment has been extensively used \cite{HH, Vilenkin, Pimentel, AdamMarosek} as much as the classical one \cite{Mariam, Alvarenga, Mariusz}.  The choice is usually made without further explanations even though the consideration of the classical scalar field keeps the theory in the semiclassical regime.  In the present work, to see the effects of both choices, we will analyzed a recently studied scenario, based on the third quantization of CQG \cite{McGuigan, Hosoya, Giddings, Strominger}, where a pair of universes are created \cite{Salva1, Salva2, Sam} in this kind of multiverse as in analogy with pair creation in quantum field theory. 

The wave function of the universe $\Psi$ is found using the Hamiltonian constraint \cite{DEWITT, Kiefer}, known as Wheeler-DeWitt equation 
\begin{equation}\label{0}
H\Psi=0,
\end{equation}
where $H$ is the Hamiltonian for a specific model.
Usually, for simplicity, while solving the Wheeler-DeWitt equation, the scalar field is kept classical. Accordingly, the degree of freedom of the scalar field $\phi$ is derived into the barotropic index $\omega$ when we consider its equation of state like $p_{\phi}=\omega\rho_{\phi}$. The Wheeler-DeWitt equation and the wave function of the universe are, then, just dependent on the scale factor once we decide the value of $\omega$.

The alternative way is to keep the scalar field as a quantum field, defining its momentum operator as ($\hbar=1$)
\begin{equation}\label{1}
p_{\phi}^2=-\frac{\partial^2}{\partial\phi^2}.
\end{equation}
Hence the Wheeler-DeWitt equation contains the explicit kinetic term of the field and its potential, and so the wave function of the universe is dependent on the scalar field and the scale factor.

For a universe with cosmological constant $\Lambda$ and a scalar field $\phi$,  the Hamiltonian reads \cite{Kiefer}
\begin{equation}\label{2}
H=\frac12\left[-\frac{p_a^2}{a}+\frac{p_{\phi}^2}{a^3}-aK+a^3\left(\frac{\Lambda}{3}+2V(\phi)\right)\right],
\end{equation}
where $a$ is the scale factor,  $V(\phi)$ is the potential of the scalar field, $K$ is the curvature index, and the canonical momenta are $p_{a}=-a\dt{a}$, and $p_{\phi}=a^3\dt{\phi}$. If the scalar field is classical, we can write the Wheeler-DeWitt equation (\ref{0}) with the Hamiltonian in Eq. (\ref{2}) by means of its density $\rho_{\phi}=\dt{\phi}^2/2+V(\phi)$ like
\begin{equation}\label{3}
\left[\frac{\partial^2}{\partial\alpha^2}-\text{e}^{4\alpha}K+\text{e}^{6\alpha}\left(\frac{\Lambda}{3}+2\rho_{\phi}(\alpha,\omega)\right)\right]\Psi_C(\alpha)=0,
\end{equation}
where we quantized the scale factor using the ordering
\begin{equation}\label{4}
p_a^2=-\frac{1}{a}\frac{\partial}{\partial a}\left(a\frac{\partial}{\partial a}\right),
\end{equation}
the density of the scalar field is expressed in terms of the scale factor as \cite{Kolb}
\begin{equation}\label{Density}
\rho_{\phi}(\alpha,\omega)=\rho_o\text{e}^{-3\alpha(1+\omega)},
\end{equation}
$\rho_o$ is the density at a certain time, and we used the parametrization $\alpha=\ln(a)$. Here, the label $C$ of the wave function stands for {\it classical}.

On the contrary, if the scalar field is quantized as in Eq. (\ref{1}), then Eq. (\ref{2}) is written like
\begin{equation}\label{5}
\left[\frac{\partial^2}{\partial\alpha^2}-\frac{\partial^2}{\partial\phi^2}-\text{e}^{4\alpha}K+\text{e}^{6\alpha}\left(\frac{\Lambda}{3}+2V(\phi)\right)\right]\Psi_Q(\alpha,\phi)=0.
\end{equation}
In this case, the label $Q$ stands for {\it quantum}.

These two treatments of the scalar field change the outcome of the differential equation. The wave functions for the classical and the quantum procedure are clearly different, and so it will be for any derived variable. 

In Section \ref{SecMath}, we will explain the formalism we use to find the entanglement entropy of a pair of universes.  Later we will analyze different models in which the entanglement entropy yields interesting outcomes.  The de-Sitter case is given in Section \ref{SecDS}, it is followed by a stiff matter dominated universe in Section \ref{SecStiff}, and finally, a closed universe with a scalar field in Section \ref{SecCU}. The conclusions are gathered in Section \ref{SecConc}.

\section{Entanglement Entropy of a Pair of Universes}\label{SecMath}

In the formalism of the third quantization which we consider,  the universes are treated like particles of a field. This way, two universes can be born together like particle and antiparticle \cite{Salva2}, as in analogy to quantum field theory. The initial ground state is $|00\rangle$, where we use the notation $|U_-U_+\rangle$ for the combined universe-antiuniverse state. During the evolution of the universes, the ground state changes in this diagonal representation, which means that the number of universes is not constant \cite{BD,Mukhanov}, in general. Since our universe seems not to change from our point of view, the most natural representation is the one in which the number of universes is invariant, called the invariant representation. It is found for systems that are analogous to the time-dependent harmonic oscillator \cite{Lewis,Vacuum}.  For a scalar field whose potential energy $V(\phi)$ is constant, the Wheeler-DeWitt equations (\ref{3}) and (\ref{5}) have the form \cite{Sam}
\begin{equation}\label{HOWDW}
\left[\pdv[2]{}{\alpha}+\omega^2(\alpha,E_{\phi},K,\Lambda)\right]\Psi(\alpha)=0,
\end{equation}
which reminds us of a time-dependent harmonic oscillator-like equation. Here, $E_{\phi}$ is the energy of the scalar field and $\omega(\alpha,E_{\phi},K,\Lambda)$ is the frequency which depends on the scale factor and the constants of the chosen model.  The invariant representation is found from the diagonal representation through the Bogoliubov transformation \cite{Bog,BD,Mukhanov} whose coefficients are \cite{Kim}
\begin{equation}
\alpha_B=\frac12\left[\frac{1}{R\sqrt{\omega}}+R\sqrt{\omega}-\frac{\i\dt{R}}{\sqrt{\omega}}\right],
\end{equation}
and
\begin{equation}
\beta_B=-\frac12\left[\frac{1}{R\sqrt{\omega}}-R\sqrt{\omega}-\frac{\i\dt{R}}{\sqrt{\omega}}\right],
\end{equation}
where\footnote{This expression can be generalized. For further details check Refs.  \cite{Vacuum} and \cite{Kim}.}
\begin{equation}
R:=\sqrt{\Psi^2_{(1)}(\alpha)+\Psi^2_{(2)}(\alpha)},
\end{equation}
and $\Psi_{(1,2)}(\alpha)$ are two real solutions of the Wheeler-DeWitt equation (\ref{HOWDW}).
The vacuum state $\ket{00}_{\text{i}}$ in the invariant representation can be expressed in terms of the diagonal states $\ket{n_-n_+}_{\text{d}}$, where $n$ labels the modes of excitation of both universes, as \cite{Mukhanov,Salva3}
\begin{equation}\label{Sum}
\ket{00}_{\text{i}}=\frac{1}{\abs{\alpha_B}}\sum_{n=0}^{\infty}\left(\frac{\abs{\beta_B}}{\abs{\alpha_B}}\right)^n\ket{n_-n_+}_{\text{d}}.
\end{equation}
Thus, from the density matrix $\rho=|00\rangle_{\text{i}}\langle00|$, we find the reduced density matrix $\rho_R$ tracing out the degrees of freedom of one of the universes in the diagonal representation.  Finally, we find the von Neumann entropy \cite{Von}
\begin{equation}\label{Ent}
S_{\text{ent}}=-\Tr{\rho_R\ln(\rho_R)},
\end{equation}
which is considered the entanglement entropy of a bipartite system like our pair.

The entanglement entropy is a measurement of the departure from the purity of the bipartite system, that is a non-entangled system. In this sense, we can identify the entanglement entropy to be recognized as a measurement of its {\it quantumness}.

As the universe increases in size, one can expect that the quantum behaviour disappears \cite{Deco}, that is, the entanglement entropy decreases. It was proved, though, that, at the critical points of the classical evolution of the universe, the entanglement entropy could behave differently, being divergent in most of the cases \cite{Sam}.

\section{de-Sitter Universe}\label{SecDS}

The de-Sitter universe \cite{Kolb} is a flat universe dominated by a cosmological constant, or a scalar field which follows the equation of state $p_{\phi}=-\rho_{\phi}$. This means that its energy is purely potential energy and $\dt{\phi}=0$. The kinetic term of the scalar field is then removed, and the potential $V(\phi)$ is found to be constant.

It is relevant to point out that the uncertainty principle constraints any quantum system. A degree of freedom like a scalar field is under the effects of this principle when it is considered quantum. The inclusion of a quantum scalar field that mimics the cosmological constant violates the uncertainty principle. However, the de-Sitter universe is always treated as a universe with a cosmological constant whose nature is not usually described. In this case, its nature is relevant. We demand the cosmological constant to be due to the potential of a scalar field whose kinetic term vanishes. As a consequence, Eqs. (\ref{3}) and (\ref{5}) become identical:
\begin{equation}\label{6}
\left[\frac{\partial^2}{\partial\alpha^2}+\text{e}^{6\alpha}\frac{\Lambda}{3}\right]\Psi(\alpha)=0,
\end{equation}
with the relation
\begin{equation}\label{7}
\Lambda=6\rho_{\phi}.
\end{equation}

The solutions to Eq. (\ref{6}), if $\Lambda\ne0$, go like (see, from here on, for special functions, Ref.  \cite{Nist})
\begin{subequations}\label{8}
\begin{gather}
\Psi_1(\alpha)\propto\mathcal{J}_0\left[\frac13\sqrt{\frac{\Lambda}{3}}\text{e}^{3\alpha}\right],
\\
\Psi_2(\alpha)\propto\mathcal{Y}_0\left[\frac13\sqrt{\frac{\Lambda}{3}}\text{e}^{3\alpha}\right],
\end{gather}
\end{subequations}
where $\mathcal{J}$ and $\mathcal{Y}$ are the Bessel functions of the first and second kind, respectively.

\begin{figure}[h]
  \centering
  \includegraphics[width=7.5cm]{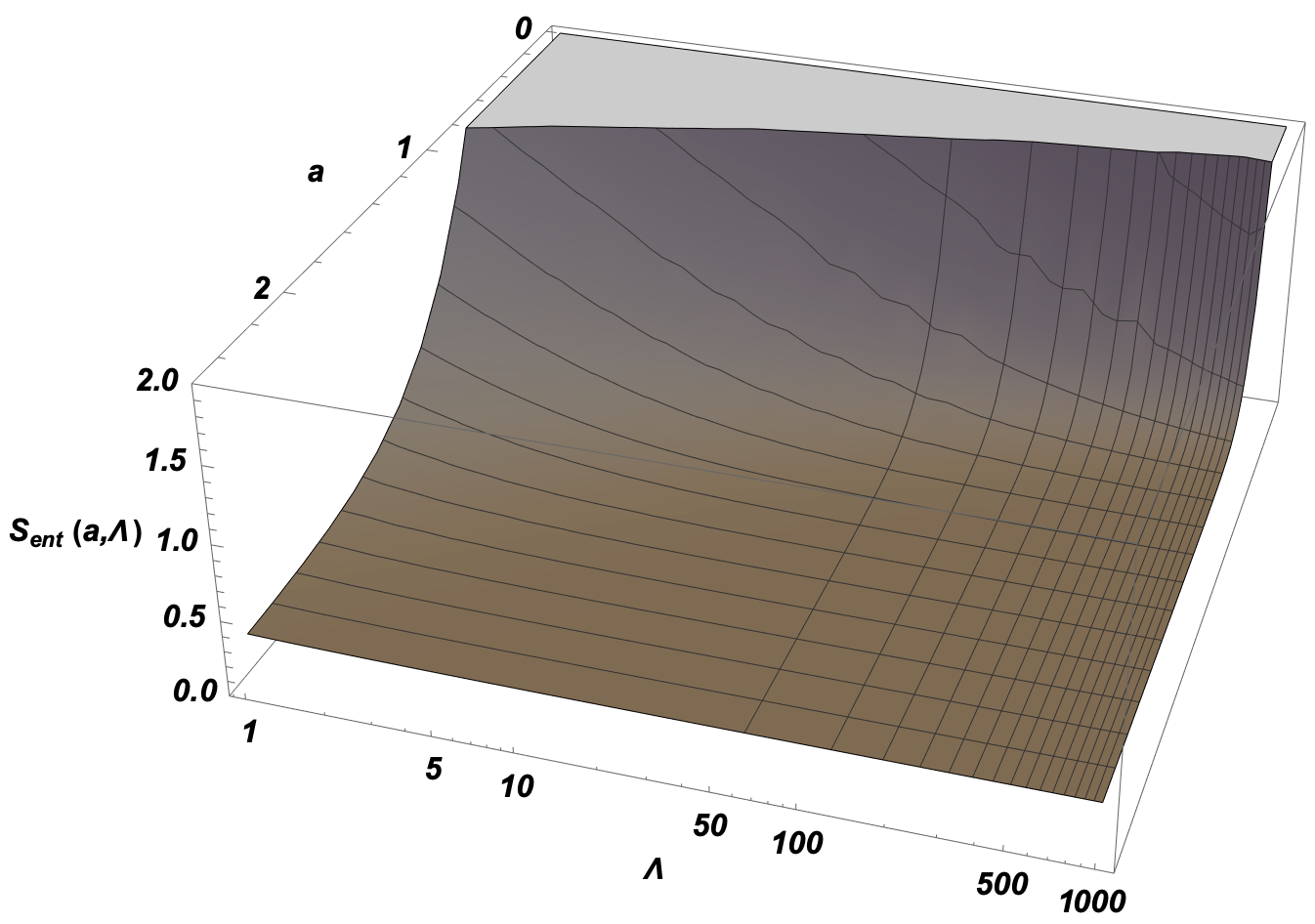}
  \caption{Entanglement entropy of a de-Sitter universe as a function of the scale factor and the cosmological constant. The entanglement entropy diverges when it goes closer to the initial singularity and it decreases as the universe gets bigger as expected by decoherence. Besides, $\Lambda$ seems to control how fast the entanglement entropy decreases.}\label{Fig1}
\end{figure}

To find the entanglement entropy, we need two real functions as solutions. Both solutions in Eqs. (\ref{8}) are real, so we use them as inputs for the calculation. The normalization constants of the solutions are not relevant here since the behaviour of the entanglement entropy is not going to change.  It can be seen by inspection of Eq. (\ref{Sum}) and how the normalization constants appear. Therefore, the results we show in this section are just qualitative. The entanglement entropy in Eq.  (\ref{Ent}), as a function of the scale factor and $\Lambda$, is depicted in Fig. \ref{Fig1}. In it, we find that the entanglement entropy is a monotonically decreasing function as the scale factor increases. It coincides with the expected vanishing of the quantumness of the system. We also see that, as the cosmological constant increases, the function is steeper. Thus, the cosmological constant seems to control the decoherence of the pair of the de-Sitter universes, being faster as $\Lambda$ gets bigger.

At the initial singularity, we find a divergent entropy. There is nothing strange about the divergent entropy since the von Neumann entropy is bounded in the interval $[0,\log(\dim(\mathcal{H}))]$, where $\mathcal{H}$ is the Hilbert space involved \cite{Information}, and the dimension of our Hilbert space $\mathcal{H}$ is infinite, that is, the invariant vacuum state of the pair $|00\rangle_i$ is written, in Eq. (\ref{Sum}), as an infinite sum of diagonal states $|n_-n_+\rangle_d$. Hence, the entanglement entropy is not bounded from above.

The de-Sitter universe does not distinguish between the quantized scalar field or the classical one. This property looks interesting when trying to analyse or work with the entanglement entropy of a system and one does not want to be mistaken due to the treatment of the scalar field.

\section{Stiff Matter Dominated Universe}\label{SecStiff}

For a stiff matter dominated universe \cite{Zel}, the dominating field satisfies the equation of state $p_{\phi}=\rho_{\phi}$. The energy of the field is, then, only kinetic, it is $V(\phi)=0$. If the scalar field is quantized as in Eq. (\ref{1}), the Wheleer-DeWitt equation from Eq. (\ref{2}) is found to be
\begin{equation}\label{9}
\left[\frac{\partial^2}{\partial\alpha^2}-\frac{\partial^2}{\partial\phi^2}\right]\Psi_Q(\alpha,\phi)=0.
\end{equation}

Following the classical procedure, the density (\ref{Density}), when $\omega=1$,  is
\begin{equation}\label{10}
\rho_{\phi}=\rho_o\text{e}^{-6\alpha}.
\end{equation}
Substitution of the density of the scalar field into Eq. (\ref{3}) yields
\begin{equation}\label{11}
\left[\frac{\partial^2}{\partial\alpha^2}+2\rho_o\right]\Psi_C(\alpha)=0.
\end{equation}

The only important difference between Eqs. (\ref{9}) and (\ref{11}) is the number of variables of the wave function of the universe. However, they essentially represent the same scenario since, here, we have two systems that are analogous to the time-independent harmonic oscillator. That means that the diagonal and the invariant representations are the same, and there will be no entanglement at all.

To check it, let us find the solutions of Eq. (\ref{9}) defining a separable ansatz
\begin{equation}\label{12}
\Psi(\alpha,\phi)=\varphi(\alpha)\chi(\phi).
\end{equation}
Thus, we divide Eq. (\ref{9}) into two different differential equations:
\begin{gather}\label{13}
\frac{\partial^2}{\partial\alpha^2}\Psi(\alpha,\phi)=E_{\alpha}\Psi(\alpha,\phi),
\\
-\frac{\partial^2}{\partial\phi^2}\Psi(\alpha,\phi)=E_{\phi}\Psi(\alpha,\phi),
\end{gather}
where $E_{\alpha}$ and $E_{\phi}$ are the energies associated to the variables, and we find that $E_{\phi}=-E_\alpha$ since $H\Psi=0$. Then, the solutions of Eq. (\ref{7}) are
\begin{equation}\label{14}
\Psi_Q(\alpha,\phi)=\text{e}^{\text{i}k_Q\left(\alpha\pm\phi\right)},
\end{equation}
where $k_Q=\sqrt{E_{\phi}}$. This is a two-dimensional plane wave, as expected.

The solutions of Eq. (\ref{11}), where the scalar field remains classical, are
\begin{equation}\label{15}
\Psi_C(\alpha)=\text{e}^{\pm\text{i}k_C\alpha},
\end{equation}
where $k_C=\sqrt{2\rho_o}$. This is another plane wave, but a one-dimensional one.

The two real solutions we need to find the entanglement entropy can be taken as the real and the imaginary part of any of the solution (\ref{14}) and (\ref{15}).  Here, we need to find the normalization constant to get the right result.  To recover the normalization, we impose on the wave function an asymptotic behavior close to the initial singularity $(\alpha\to-\infty)$,  where only incoming positive frequency plane waves of the form \cite{mukhanov_winitzki_2007}
\begin{equation}\label{BC}
\Psi(\alpha)=\frac{1}{\sqrt{k}}\e^{-\i k \alpha},
\end{equation}
are allowed.  Thus, the entanglement entropy (\ref{Ent}) is found to be the zero function.  Without the right normalization, the result is a constant entropy.  The vanishing of the entanglement entropy  proves that both representations are the same.  The stiff matter dominated universe seems to suffer the lack of quantum correlations with its twin universe. This kind of universe looks like the most natural system to avoid entanglement in the multiverse.

\section{Closed Universe with a Scalar Field}\label{SecCU}

To finish with our analysis, we consider a non-trivial system. In this case, we have a scalar field in a closed universe $(K=1)$.  The equation of state of the scalar field is not known in general. If we consider it to be a perfect fluid, we can be safe studying an interval for the barotropic index $\omega\in[-1,1]$, since
\begin{equation}
\omega=\frac{\dt{\phi}^2/2-V(\phi)}{\dt{\phi}^2/2+V(\phi)},
\end{equation}
and we assume that the kinetic energy of the field is always positive.

Keeping the scalar field as classical as in Eq. (\ref{3}), it yields the Wheeler-DeWitt equation
\begin{equation}\label{19}
\left[\frac{\partial^2}{\partial\alpha^2}-\text{e}^{4\alpha}+2\rho_o\text{e}^{3\alpha(1-\omega)}\right]\Psi_C(\alpha)=0,
\end{equation}
whose solutions are not easily found. We performed a numerical calculation, with the boundary condition in Eq. (\ref{BC}), to find the solutions in terms of $\alpha$ and $\omega$. 

\begin{figure}[h]
  \centering
  \includegraphics[width=7.5cm]{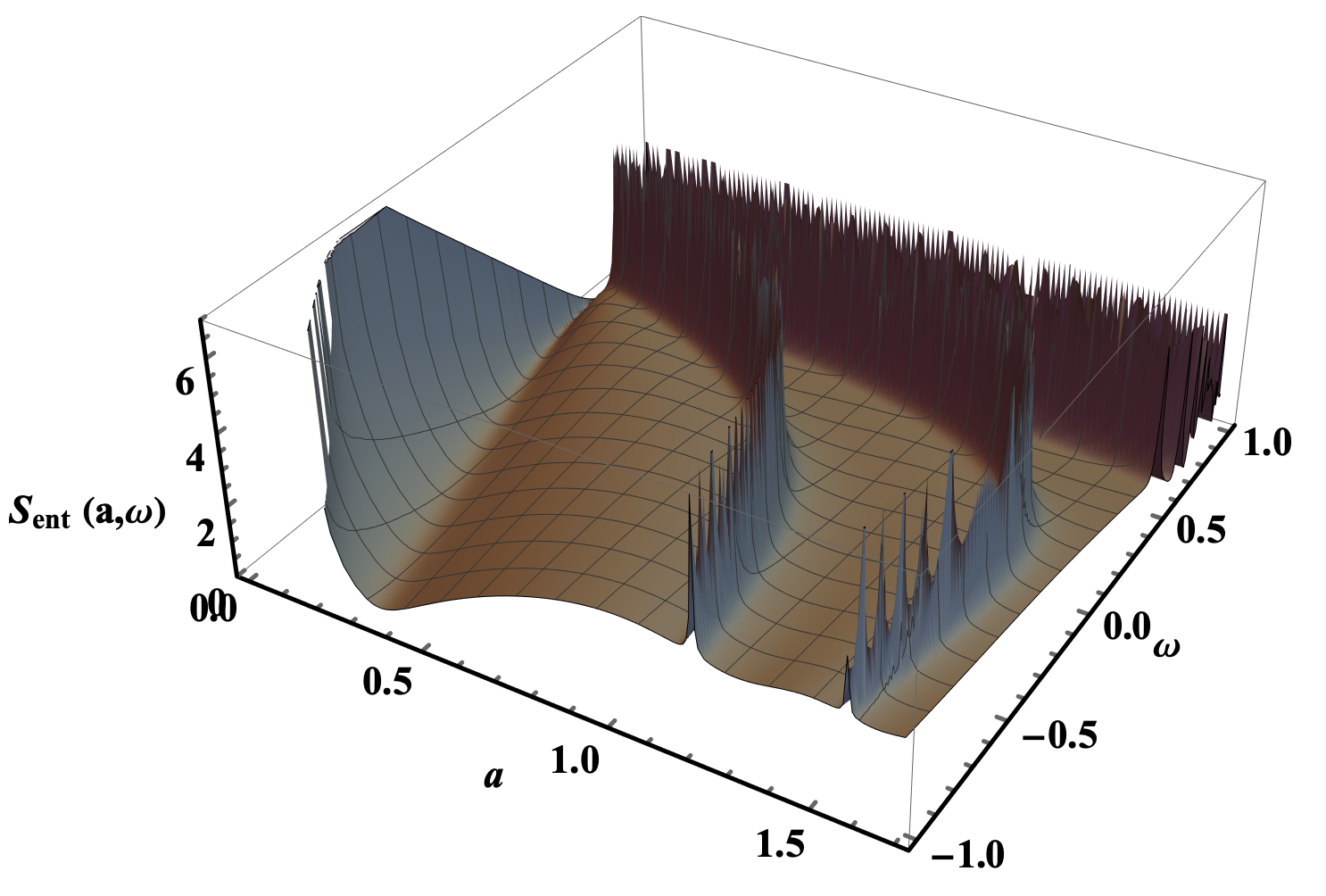}
  \caption{Entanglement entropy of a closed universe with a scalar field considered classical, as a function of the scale factor and the barotropic index $\omega$. Here, we used $\rho_o=1/2$. We see that there is a decreasing behavior for the entanglement entropy when the scale factor increases. The vertical lines are due to the numerical method and the derivatives of the fast oscillating wave functions, and should not be taken into account as part of the real entanglement entropy.} \label{Fig2}
\end{figure}

Using the real part and the imaginary part of a solution, the entanglement entropy (\ref{Ent}) is calculated. It is shown, with $\rho_o=1/2$, in Fig. \ref{Fig2}. The figure is a bit noisy due to the numerical method and the derivatives of the fast oscillations of the wave functions that are needed. One can see that, as we approach the initial singularity, it increases rapidly, and it happens for any value of $\omega$. As the universe gets bigger, the entanglement entropy starts to oscillate slowly. How the entropy behaves for high values of $a$ is hard to say. The frequencies of the wave functions are increasing with the scale factor, and it makes the numerical method to break down. However, we only need the behaviour close to the singularity to see the difference with the entanglement entropy due to the system with the quantized scalar field.

\begin{figure}[h]
  \centering
  \includegraphics[width=8cm]{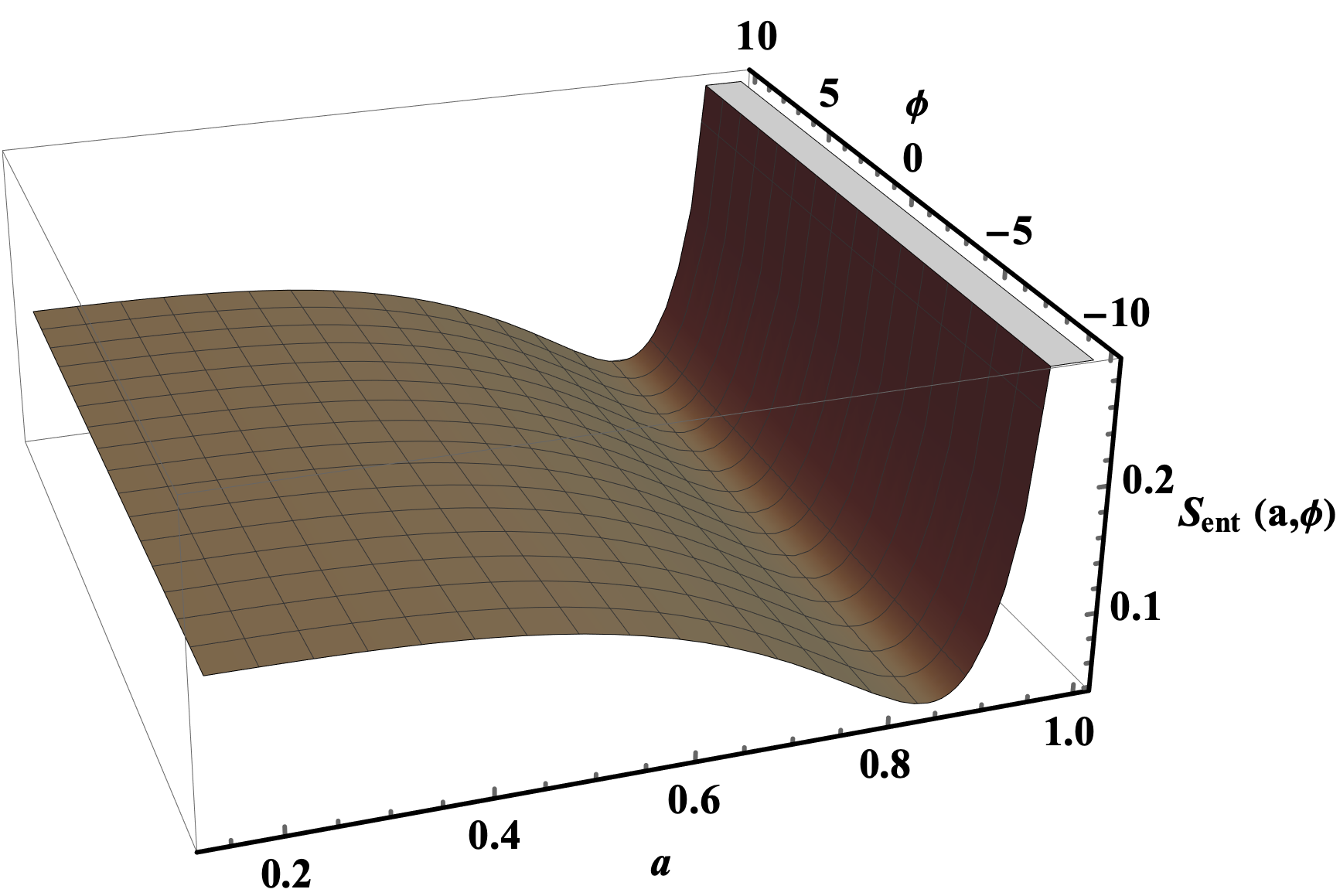}
  \caption{Entanglement entropy of a closed universe with a quantized scalar field. It is represented the entanglement entropy of a universe with a scalar field whose potential is vanishing. The divergence at $a=1$ appears because it is the maximum size of the universe. At the initial singularity, the entropy is finite and decreasing.}\label{Fig3}
\end{figure}

Now, for a quantized scalar field, we inspect Eq. (\ref{5}) including any potential $V(\phi)$:
\begin{equation}\label{WdW}
\left[\frac{\partial^2}{\partial\alpha^2}-\frac{\partial^2}{\partial\phi^2}
-\text{e}^{4\alpha}+2V(\phi)\text{e}^{6\alpha}\right]\Psi_Q(\alpha,\phi)=0.
\end{equation}
It has not simple solutions for a non-vanishing potential, in general. Our point here, is to analyse only the asymptotic behavior, close to the initial singularity, of the entanglement entropy. That means that the last term of Eq. (\ref{WdW}), the one which contains the potential of the scalar field, can be neglected compared with the third term.  The differential equation is separable in that case. With the ansatz
(\ref{12}), one finds
\begin{equation}
\chi(\phi)=\text{e}^{\pm\i \sqrt{E_{\phi}} \phi},
\end{equation}
where $E_{\phi}$ is the energy associated to $\phi$, and
\begin{equation}
  \varphi(\alpha)=\mathcal{I}_{\pm\text{i}\frac{\sqrt{E_{\phi}}}{2}}\left[\frac12\text{e}^{2\alpha}\right],
\end{equation}
where $\mathcal{I}_{\nu}(z)$ is the modified Bessel function of the first kind. Thus, the solutions of Eq. (\ref{WdW}) are
\begin{equation}
  \Psi(\alpha,\phi)=\mathcal{I}_{\pm\text{i}\frac{\sqrt{E_{\phi}}}{2}}\left[\frac12\text{e}^{2\alpha}\right]\text{e}^{\pm\i \sqrt{E_{\phi}} \phi}.
\end{equation}
The real and the imaginary part of any of those solutions can be used as the inputs for the calculation of the entanglement entropy. We found and displayed it, using $E_{\phi}=1$, in Fig. \ref{Fig3}. It is clear that this entropy is not the associated to the pair of universes following Eq. (\ref{WdW}) unless the potential is null. There, we see that the entropy decreases as the universe gets bigger, which coincides with our expectations. However, when it gets closer to $a=1$, it diverges since it is a critical point \cite{Sam} of the universe that follows Eq. (\ref{WdW}) with $V(\phi)=0$ along its entire evolution.

But for a pair of universes described by Eq. (\ref{WdW}), the entanglement entropy shown in Fig. \ref{Fig3} is only correct when $\alpha\to-\infty$. The difference between Fig. \ref{Fig2} and Fig. \ref{Fig3}, when we look close to the initial singularity, is that the entanglement entropy diverges in the former case, and it is finite in the latter case. For us to prove that there is a difference when we treat the scalar field differently, it is enough.

\section{Conclusions}\label{SecConc}

The importance of considering a scalar field as classical or quantum has been analyzed.  In order to check how it makes a difference, we have calculated the entanglement entropy of a pair of universes in both cases for three different models: de-Sitter universes, flat stiff matter dominated universes, or a close universe with a general scalar field.

For the de-Sitter universes, we found, independently of the treatment of the scalar field, that the entanglement entropy is unchanged. The pair of de-Sitter universes is found to be a good option not to make any mistake when treating the scalar field as classical. 

The trivial case is the stiff matter dominated universe, where the entanglement entropy vanishes along the entire evolution, and for any value of the scalar field. The reason is that the diagonal and the invariant representations are the same in this scenario.

Those two first cases demonstrate that there could be systems for which both considerations are irrelevant. However, in the last case where we considered a closed universe with a scalar field, we can see what the effect of the treatment of the scalar field on the entanglement entropy is comparing Fig. \ref{Fig2} with Fig. \ref{Fig3}. To make such comparison, we must focus our attention at the smallest values of the scale factor, because of the restriction we made in Section \ref{SecCU} simplifying Eq. (\ref{WdW}).  For the universes where the scalar field is considered as classical, it diverges rapidly.  On the other hand, it is finite when we keep a quantum scalar field.  It proves that anything we get using different treatments of the scalar field will yield, perhaps, different results. It is evident since considering a classical scalar field forces the theory into a semiclassical theory of gravity.

Finally, we would like to point out some other interesting ideas. If one imagines the universe to start completely empty except for a cosmological constant, given by the vacuum energy density or any other reason, that is, a de-Sitter universe, the entanglement between the pair of universes exists and it is,  at the very beginning, of the form we found in Section \ref{SecDS}.

Furthermore, after the result found in Section \ref{SecStiff} for the stiff matter dominated universe, where there is no entanglement at all, the comparison with other methodologies (see e.g. \cite{Kanno, Holman}) could be an interesting avenue for further research, since it seems to be a fast way to check the mutual consistency of the models.

\section*{Acknowledgments}
The author would like to thanks F.  Wagner for pointing out some details to comment. The work was supported by the Polish National Research and Development Center (NCBR) project ''UNIWERSYTET 2.0. --  STREFA KARIERY'', POWR.03.05.00-00-Z064/17-00 (2018-2022).

\bibliography{Sam.bib}

\end{document}